# Microbiome and metabolome insights into the role of the gastrointestinal-brain axis in neurodegenerative diseases: unveiling potential therapeutic targets

Helena U. Zacharias[1,2,3,*,#], Christoph Kaleta[4,5,*,#], Francois Cossais[6], Eva Schaeffer[7], Henry Berndt[9], Lena Best[4], Thomas Dost[4], Svea Glüsing[10], Mathieu Groussin[3], Mathilde Poyet[11], Sebastian Heinzel[7,12], Corinna Bang[3], Leonard Siebert[13,5], Tobias Demetrowitsch[10,14], Frank Leypoldt[7,8], Rainer Adelung[13,5], Thorsten Bartsch[7,5], Anja Bosy-Westphal[15], Karin Schwarz[10,5], Daniela Berg[7,5]

[1]Peter L. Reichertz Institute for Medical Informatics of TU Braunschweig and Hannover Medical School, Hannover, Germany.
[2]Department of Internal Medicine I, University Medical Center Schleswig-Holstein, Campus Kiel, Kiel, Germany.
[3]Institute of Clinical Molecular Biology, Kiel University and University Medical Center Schleswig-Holstein, Campus Kiel, Kiel, Germany.
[4]Research Group Medical Systems Biology, Institute for Experimental Medicine, Kiel University and University Medical Center Schleswig-Holstein, Kiel, Germany.
[5]Kiel Nano, Surface and Interface Science - KiNSIS, Kiel University, Kiel, Germany.
[6]Institute of Anatomy, Kiel University, Kiel, Germany.
[7]Department of Neurology, Kiel University and University Medical Center Schleswig-Holstein Campus Kiel, Germany.
[8]Neuroimmunology, Institute of Clinical Chemistry, Kiel University and University Medical Center Schleswig-Holstein, Kiel, Germany.
[9]Research Group Comparative Immunobiology, Zoological Institute, Kiel University, Kiel, Germany.
[10]Institute of Human Nutrition and Food Science, Food Technology, Kiel University, Kiel, Germany.
[11]Department of Biological Engineering, Massachusetts Institute of Technology, Cambridge, MA, USA.
[12]Institute of Medical Informatics and Statistics, Kiel University, University Hospital Schleswig-Holstein, Kiel, Germany.
[13]Functional Nanomaterials, Department of Materials Science, Kiel University, Kiel, Germany.
[14]Kiel Network of Analytical Spectroscopy and Mass Spectrometry, Kiel University, Kiel, Germany.
[15]Institute of Human Nutrition and Food Science, Kiel University, Kiel, Germany.

*These authors contributed equally to this work.
[#]Correspondence: helena.zacharias@plri.de and c.kaleta@iem.uni-kiel.de.

## Abstract

Due to the aging of the world population and westernization of lifestyles, the prevalence of neurodegenerative diseases such as Alzheimer's disease (AD) and Parkinson's disease (PD) is rapidly rising and is expected to put a strong socioeconomic burden on health systems worldwide. Due to the limited success of clinical trials of therapies against neurodegenerative diseases, research has extended its scope to a systems medicine point of view, with a particular focus on the gastrointestinal-brain axis as a potential main actor in disease development and progression. Microbiome as well as metabolome studies along the gastrointestinal-brain axis have already revealed important insights into disease pathomechanisms. Both the microbiome and metabolome can be easily manipulated by dietary and lifestyle interventions, and might thus offer novel, readily available therapeutic options to prevent the onset as well as the progression of PD and AD. This review summarizes our current knowledge on the association between microbiota, metabolites, and neurodegeneration in light of the gastrointestinal-brain axis. In this context, we also illustrate state-of-the art methods of microbiome and metabolome research as well as metabolic modeling that facilitate the identification of disease pathomechanisms. We conclude our review with therapeutic options to modulate microbiome composition to prevent or delay neurodegeneration and illustrate potential future research directions to fight PD and AD.

## The gastrointestinal-brain axis as a potential mediator of microbiome effects in neurodegenerative diseases

In the last few years there has been a growing understanding of pathophysiological cascades and molecular changes involved in the manifestation of neurodegenerative diseases like Parkinson's (PD) and Alzheimer's disease (AD). However, triggering factors initiating these cascades, modulating factors influencing disease progression as well as early interventional approaches addressing these factors remain elusive. Although monogenic forms of both PD and AD are known, the majority of cases are idiopathic with complex and heterogeneous etiological contributions from a multitude of possible genetic and/or environmental risk factors. Lifestyle factors like physical activity and diet, in particular, may very well constitute modifiable risk factors of AD and PD manifestation and progression [1,2]. Elucidating these factors is highly needed due to the enormous increase in PD and AD prevalence, which exceeds the increase that can be expected from an aging world population alone [3,4].

The nervous system of the GI tract, which contains 200-500 million neurons, is in close exchange with the central nervous system (CNS). This bidirectional communication is often referred to as the gut-brain axis. However, as it also involves the upper GI tract, including the mouth and its specific microbial environment, we hereafter use the broader term "GI-brain axis". Several modes of communication along the GI-brain axis have been described, which can be summarized as neurochemical, endocrine and immune interaction [5]. Yet, the breadth of mechanisms involved in this communication are only poorly understood. A growing body of research now suggests that our microbiota, the diverse and complex communities of commensal microbes that colonize all our body surface barriers, play a key role in the GI-brain axis, and may be involved in neurodegenerative diseases. Closely interconnected with the microbiome is the metabolome, the complete set of small molecules, called metabolites, which are intermediate or end-products of metabolism. Their involvement in neurodegenerative diseases currently also attracts wide interest in the research community.

In this review, we summarize the current knowledge on the association between microbial imbalance and neurodegeneration. Furthermore, we review state-of-the art association studies between neurodegeneration and the metabolome in PD and AD, the role of metabolic modeling in defining molecular pathways underlying those associations, as well as the potential of both the microbiome and the metabolome as novel therapeutic targets to treat neurodegeneration [6].

## From the microbiome to the metabolome

Human microbiota are mostly composed of bacteria, but also contain archaea and microbial eukaryotes, along with their associated viral communities. The collection of genes encoded by these microbial communities defines the microbiome. Microorganisms produce large quantities of diverse molecular compounds that directly influence host metabolism, prime immune responses and shape physiology [7]. They also harbor complex surface markers that engage in direct contact interaction with host receptors or circulating proteins, which can either trigger anti- or pro-inflammatory responses [7]. As such, endogenous microbes are suspected to be strong contributors to our health via constant inter-organ interactions, also including the CNS [8]. However, detailed and mechanistic knowledge about these signaling pathways is limited, even though recent progress has been made [9]. In particular, the most abundant of all microbial metabolites are the short-chain fatty acids (SCFAs) acetate, propionate and butyrate [10], produced from the metabolism of indigestible materials, such as complex fibers. Their downstream effects on host physiology are very diverse: while acetate is readily absorbed into the bloodstream and distributed to peripheral tissues, propionate is metabolized by the liver after absorption [11]. The majority of butyrate, on the other hand, is consumed locally by colonocytes as a primary fuel source [12]. Importantly, SCFAs have been shown to contribute to preventing pathogen invasion and to shaping the immune system [7].

***Interrogating the microbiome***

Historically, culture-independent methods were used to characterize the diversity and structure of the microbiota by amplicon sequencing of phylogenetic marker genes (e.g. the 16S rRNA gene) [13]. Amplicon-based data are usually restricted in taxonomic, genomic and functional information, limiting our understanding of the differences in microbial features that may exist between two given microbiomes. Today, deep shotgun metagenomic sequencing is used as the gold-standard approach to interrogate microbiomes [14,15]. Combined with sophisticated computational methods that reconstruct draft genomes [16], identify microbial lineages at the resolution of strains [17,18], or reconstruct gene repertoires [19] with detailed functional annotations [20], metagenomics data provide high-dimensional and complex data that, when used alone or integrated with other multi-omics data, can reveal insightful associations between the microbiome and disease phenotypes [21,22].

Microbiome signatures of disease usually exhibit a loss of taxonomic diversity, decrease in the abundance of microbes that are suspected to be beneficial, and increase in abundance of potential pathobionts [23]. However, microbiome association studies have suffered from a lack of replicability across different cohorts [23] concerning the identification of microbiome features that are associated with disease. Both biological and methodological aspects contribute to this problem. We now know that inter-individual variability is high in human microbiome data [24], and that geography can have a larger effect on microbiome variance than any other disease-relevant human trait, such as drugs, diet or genetics [25,26]. Therefore, large sample sizes, appropriate geographic (and/or lifestyle) representation and extensive surveys of metadata are needed to limit the effect of confounders and to draw reliable conclusions [21]. Importantly, technical protocols (e.g. to extract microbial DNA) and the choice of experimental or computational tools can explain more variance in microbiome sequencing results than single host traits [27,28]. Finally, it is now commonly expected that employed methods can account for the specific properties of microbiome data, in order to increase specificity and sensitivity of association analyses and promote cross-study comparisons [23,29]. Particularly, microbiome data, which are count data, are compositional (quantification data for each taxa are usually in relative, and not absolute, abundance), high-dimensional (hundreds or thousands of microbial taxa are detected among a given set of samples), sparse (a large fraction of microbial taxa are detectable only in a subset of samples), and overdispersed (variances of count data are larger than would be expected under a Poisson model) [29]. Overall, this illustrates the need to embrace experimental and computational standards among the microbiome research community to favor reproducibility and cross-study meta-analyses [30].

***Microbiome signatures of disease***

Both AD and PD differences in gut microbiome features compared to healthy controls have been observed [31,32]. It has been speculated that differences in the taxonomic diversity and composition of the GI tract microbiota results in perturbations of metabolic and immune-microbe interactions, thereby contributing to disease pathology. The causal role of a disturbed microbial homeostasis on pathogenesis is supported by fecal microbiota transplant (FMT) experiments in which disease phenotypes could be transferred from affected individuals to germ-free recipient animals through microbiome transfers alone. Thus, it has been postulated that dysbiotic changes of the microbiome are an important contributor to diseases such as inflammatory bowel disease (IBD) [33], PD [34], AD [35] and aging per se [36]. However, findings of FMT studies that used germ-free animals have been called into question due to the surprisingly high success rate of microbiome transfer experiments [37]. As such, mechanisms through which dysbiotic changes of the microbiome could contribute to disease processes in the host, and neurodegeneration in particular, are still poorly understood. A disbalanced microbiome could be characterized by an overabundance of pathogenic bacteria that are capable of releasing molecules, such as endotoxins, that may induce inflammation and compromise barrier integrity. Alternatively, perturbations in the quantity, or balance, of SCFAs are also suspected to be involved, as SCFAs were shown to be key microbial mediators in the GI-brain axis [9].

## Interrogating the metabolome

The comprehensive study of the metabolome in a particular biospecimen is the core goal of metabolomics (Fig. 1) [38]. As the metabolome rapidly responds to both endo- and exogenous stimuli, metabolomics can provide a metabolic “snapshot” or “fingerprint” of the current state of an organism. It is thus able to offer new insights into the pathomechanisms underlying human diseases and identify potential therapeutic targets. Metabolomics studies can be conducted either in an untargeted or a targeted manner. Untargeted metabolomics tries to maximize the metabolome coverage of an investigated biospecimen, without any a priori metabolite selection. In contrast, targeted metabolomics measures a predefined set of metabolites, and often provides absolute quantification of their concentrations.

Metabolomics measurements in biospecimens are typically conducted by nuclear magnetic resonance (NMR) spectroscopy or mass spectrometry (MS) (Fig. 1). NMR spectroscopy separates different metabolite signals according to their resonance frequencies within a magnetic field. MS, in contrast, identifies different metabolites by analyzing their mass-to-charge ratios. Considering the complex nature of biological samples, the majority of MS analysis methods involve prior analyte separation. Hyphenated techniques combine, e.g., liquid (LC) or gas chromatography (GC) with mass spectrometers. Compounds that are adequately volatile can be easily analyzed by GC-MS. In GC-MS, the electron impact ionization source allows neutral molecules to be ionized using an electron beam, and instantaneously fragments the entering molecules into a characteristic pattern [39–41]. LC-MS generally uses soft ionization techniques that mainly display the molecular ion species with only a few fragment ions. To overcome the resulting problem of rather poor information obtained from a single LC-MS run, tandem mass spectrometry (MS/MS) can be used. LC-MS/MS provides fragments through collision-induced dissociation of the molecular ions produced [42]. When compared to LC-MS based methods, GC-MS has advantages of a greater chromatographic resolution, a good retention of small compounds and large spectral libraries [43,44]. However, the thermal stability of samples limits the metabolome coverage by GC-MS. Furthermore, several metabolites require derivatization, which might produce artifacts [44].

NMR spectroscopy requires very little sample preparation and measurements are highly reproducible both across time and across different lab facilities [45]. Likewise, NMR-based metabolomics is rather cheap, since only one internal standard is required to extract absolutely quantified metabolite concentrations from spectral data. This also facilitates absolute metabolite quantification in untargeted NMR metabolomics experiments. Furthermore, NMR experiments are non-destructive, and biospecimens can be re-used afterwards. NMR spectroscopy is, however, a rather insensitive analytical technique in comparison to MS, and acquired spectral data, especially from one-dimensional NMR experiments, are highly complex. Therefore, metabolite identification and absolute quantification are, to date, still very time-consuming processes, mainly conducted manually and not yet completely automatized. High spectral metabolite signal overlap can be partially compensated by two-dimensional NMR experiments, but at the expense of significantly increased measurement times. Nevertheless, these experiments can also provide further important structural metabolite information, which can improve the identification and characterization of unknown compounds.

MS-based approaches have the advantage of high sensitivity and selectivity, as well as high throughput and depth of coverage. The applicability of direct injection in metabolomics is extended by advanced instrumentation capable of high-resolution, accurate mass measurements and tandem MS [46]**.** Fourier transform - ion cyclotron resonance mass spectrometers (FT-ICR-MS; Fig. 1) are the most advanced mass analysers in terms of information content and resolving power, with sub-parts-per-million mass accuracy [47] the possibility of direct infusion mass spectrometry, which generates data in only a few minutes.

The most comprehensive coverage of the metabolome can only be achieved by a combination of different analytical techniques, e.g., NMR and different MS methods. Therefore, these techniques should be seen as complementary rather than as competing. For detailed information on conducting metabolomics experiments, we refer the interested reader to [38,48–50].

Most metabolome analyses to identify biomarkers for AD and PD are based on cerebrospinal fluid (CSF) [51] and blood specimens [52], including plasma [53] and serum [54]. As CSF has a more immediate connection to the brain than any other fluid, it directly reflects its metabolic changes [55]. The collection of blood samples is less invasive, acceptable for repeated measures and most closely connected to CSF. Some studies have examined other biological matrices such as urine [54], feces [56], brain tissue [57], saliva [58] or sebum [59]. Urine is of great interest for biomarker identification, as it contains most of the body's metabolic end products [60] and is therefore able to reflect comprehensive changes of metabolites in organisms [61]. In addition, urine represents a non-invasive biospecimen source. The fecal metabolome is also of particular interest since it more directly captures the complex interactions between the gut microbiome and the host [62].

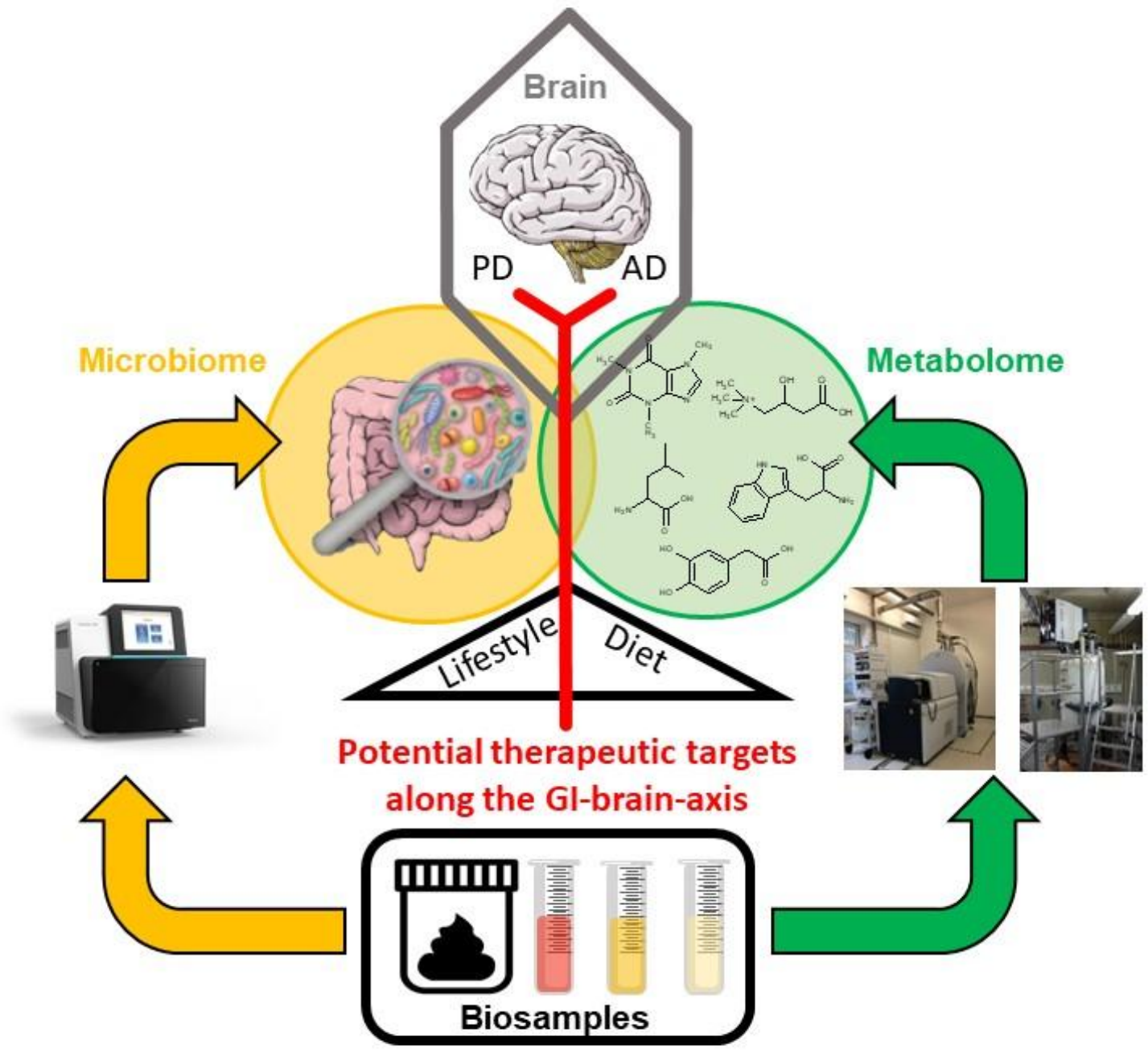

**Fig. 1** Schematic illustration of the gastrointestinal (GI)-brain axis (red) that could be modulated via potential therapeutic targets involving indirect (lifestyle/diet) and direct modulation of the microbiome (orange circle) and metabolome (green circle). Such therapeutic modulation of the GI-brain axis may represent a promising strategy for the early prevention of neurodegenerative processes in Alzheimer's and Parkinson's disease. Further research and analyses of biosamples regarding the microbiome and metabolome are needed and facilitated by methodological advances: Sequencing allows the taxonomic and functional characterization of the microbiome (orange arrows) of stool samples of the gut and biospecimen from various other bodys sites.  Metabolomics (green arrows) is enabled by mass spectrometry (left) or nuclear magnetic resonance spectroscopy (right).

## Microbiome and microbiome-linked metabolome changes in neurodegeneration

To date, most studies on PD-associated microbiota are based on 16S RNA-sequencing. Although many studies have identified alterations of fecal microbial diversity in PD, a high variability is observed in currently published datasets. Recently, several meta-analyses have attempted to provide a more unified view on microbial alterations occurring in PD [34,63]. The study by Plassais et al. indicated that microbial alpha-diversity is not significantly altered in patients with manifest PD or multiple sclerosis in comparison to healthy controls [63]. However, a second meta-analysis, showed increased alpha-diversity in PD compared to controls, and suggested a link between disease and changes in the abundance of bacterial species, as well as intestinal inflammation [34]. These changes include an enrichment of the genera *Lactobacillus*, *Akkermansia*, and *Bifidobacterium*, as well as a reduction of the Lachnospiraceae family and *Faecalibacterium* genus, which both have been described as SCFA producers. These results were partly confirmed in a subsequent analysis indicating that increased *Akkermansia* and reduced *Roseburia* are consistently found in PD [64].

Importantly, similar microbial changes in the abundance of SCFA producing taxa have also been reported for prodromal stages and markers of PD [65,66].

Microbiota alterations in PD have also been shown using shotgun metagenomic sequencing, indicating that the frequency of a subset of bacterial genes may allow to distinguish PD patients from healthy subjects, as well as from patients with multiple system atrophy (MSA) or AD [67]. Moreover, a functional metagenomic analysis suggested differences in the metabolism of SCFA precursors in PD compared to controls [68].

Noteworthy, most studies investigating a potential link between fecal microbiome and PD have suggested that microbiota-derived SCFA production may be altered in PD [69]. In particular, a reduction of fecal SCFA concentrations has been observed in PD patients [70] while the investigation of SCFA levels in serum or plasma have led to more conflicting results. A first study indicated that SCFA concentrations appear not to be significantly changed in the serum of PD patients [71] and an additional study indicated that serum SCFAs may help to distinguish between MSA and PD patients, but not between PD patients and control subjects [71,72]. To the contrary, more recent studies have reported increased plasma SCFA concentrations, particularly for acetate and propionate, in PD patients in comparison to control subjects [73]. Limited availability of SCFAs in the blood, due to their fast metabolization along the GI tract may explain these discrepancies in part.

In AD, gut dysbiosis potentially triggers increased systemic inflammation, which in turn increases penetrability of the gut mucus barrier and leads to a more transmissible blood-brain-barrier. Thus, microbiome derived metabolites, like LPS (astrocyte activation), SCFAs (anti-neuroinflammatory), secondary bile acids (neurodegenerative) or tryptophan-related metabolites (neuroinflammatory), are more likely to reach the brain. A detailed review by Bairamian and colleagues [74] has covered the role of microbiota in AD more exhaustively. AD-associated dysbiosis comprises an increase of pro-inflammatory microbes and decrease of anti-inflammatory commensals. Bacteria of the Firmicutes phylum (including butyrate producers) are reduced in aged individuals [75] as well as in AD patients [76] and furthermore in the AD mouse models 5xFAD [77] and P301L [78]. SCFA supplementation rescued an immature microglia phenotype in germ free mice [79]. Butyrate has been shown to alter microglia states towards the homeostatic M0 type [80]. Microglia are brain resident immune cells which act neuroprotective in the homeostatic state (M0) but can act pro-inflammatory in DAM or MGnD state [74]. Amyloid accumulation in aged humans with or without dementia was negatively correlated with butyrate and anti-inflammatory IL-10, whereas acetate, valerate and proinflammatory cytokines have been positively correlated [81]. In the APP/PS1 mouse model a fiber rich diet increased abundance of butyrate producing taxa, which led to reduced astrocyte activation and improved cognitive function, while propionate showed deleterious effects [82]. APOE, with the allele e4 being the biggest genetic risk factor for AD, is involved in the shift of microglia states from homeostatic (M0) to disease-associated microglia (DAM) via expression of the neuroinflammation-associated TREM2 gene [83]. Interestingly, APOE4 carriers have been shown to have reduced levels of *Ruminococcaceae*, known butyrate producers, compared to the APOE2/E3 genotype [84], and loss of these bacteria was also observed in AD patients [76]. It is currently hypothesized that microbiome derived amyloid proteins (e.g. curli) could induce Amyloid-beta-aggregation by acting as a seed [85]. Additionally, it has been shown that it is possible for α-synuclein to shuttle from gut to brain via the vagus nerve [86]. Microbial amyloid proteins might be able to take a similar route [32].

***Metabolomic changes in neurodegeneration***

Several studies already investigated metabolome changes associated with PD (Table 1). Shao et al. identified several metabolites such as caffeine metabolites and fatty acids that were significantly decreased in plasma of PD patients compared to healthy controls [87]. Hatano et al. reported caffeine-related metabolites and purine derivatives as significantly decreased only during the initial stages of PD in the serum of PD patients [88]. Furthermore, increased levels of branched-chain amino acids (BCAAs) were found in patients with PD [89]. In urine, the levels of leucine and isoleucine were positively correlated with disease stage in idiopathic PD patients [89]. In line with this finding, milk consumption (but not fermented milk intake) was associated with increased risk of PD [90,91]. Other authors however found a negative correlation between plasma BCAAs or essential amino acids (EAA) and Parkinson's disease scores [92]. In these patients, whey protein supplementation increased plasma BCAAs and EAA and led to an increase in plasma reduced glutathione and a reduction in homocysteine levels at unchanged UPDRS. Furthermore, a dysregulation of metabolites associated with carnitine metabolism was observed in serum [59] and plasma [93]. Carnitine-dependent oxidation of fatty acids is an alternative way of energy production in mitochondria. Therefore, a disturbance of the carnitine metabolic pathway could be related to the mitochondrial dysfunction observed in PD [93]. Significantly lower plasma or serum levels of tryptophan and kynurenine were reported for PD patients, indicating an involvement of this particular pathway in PD pathogenesis [53]. In a rotenone-induced rat model of PD, dietary tryptophan supplementation was shown to protect against rotenone-induced neurotoxicity to ameliorate motor deficits, which may be mediated through activating the aromatic hydrocarbon receptor pathway [94]. Similarly, metabolic profiling of whole blood samples showed increased levels of leucine in de novo PD patients compared to controls as well as higher levels of tryptophan metabolites, including kynurenine and xanthurenic acid, in PD patients compared to controls [52].

Recent studies on metabolomics in AD are summarized in Table 1. Similar to PD, alterations in serum acylcarnitine composition have been reported in incident AD and associated with cognitive decline [95]. Analysis of feces specimens revealed higher ammonia and lactic acid concentrations in subjects with dementia [96]. Targeted metabolomics analyses in serum and brain tissue demonstrated alterations of bile acid metabolism in AD, resulting in a higher proportion of secondary bile acids in comparison to healthy subjects [97–99]. Bile acids are considered important endocrine and paracrine effectors, directly linking liver homeostasis and intestinal co-metabolism with the CNS. Multiple studies investigated how bile acids cross the blood brain barrier and how they are involved in signaling circuits, emphasizing the role of the GI-(liver)-brain axis in AD [100–102]. Intestinal abundance of the genus *Faecalibacterium* correlated negatively with disease severity in dementia which was confirmed in further studies in AD and PD [69,103,104]. The role of *Faecalibacterium* as an important butyrate fermenter with anti-inflammatory effect has already been discussed for IBD [105]. A multivariable, blood-based metabolite panel might be promising to differentiate AD patients from controls and other types of dementias [106].

**Table 1: Recent metabolomics studies in (a) PD and (b) AD.** Abbr.: AD, Alzheimer's disease; BCAA, branched chain amino acid; BMI; body-mass-index; CE-FTMS, capillary

electrophoresis Fourier transform mass spectrometry; CSF, cerebrospinal fluid; FT-ICR-MS; Fourier transform ion cyclotron resonance mass spectrometry; GC-MS; gas chromatography mass spectrometry; LC-MS, liquid chromatography mass spectrometry; LC-MS/MS, liquid chromatography tandem mass spectrometry; MCI, mild cognitive impairment; MRI, magnetic resonance imaging; NMR, nuclear magnetic resonance; PET, positron emission tomography; PD, Parkinson's disease; SCFA, short-chain fatty acid; TMAO, trimethylamine-N-oxide.

| **Publication** | **Study Question** | **Analytical Method** | **Sample Matrix** | **Additional measurements** | **Study population** | **Finding (detected metabolites/metabolic biomarkers/pathways)** |
|---|---|---|---|---|---|---|
| **(a) Parkinson's Disease** | | | | | | |
| [73] | compare fecal and plasma levels of different SCFA subtypes in patients with PD and healthy controls | GC-MS and LC-MS/MS | feces and plasma | total fecal DNA | 96 PD patients and 85 controls | reduced fecal SCFAs and increased plasma SCFAs observed in patients with PD and correlated to specific gut microbiota changes and clinical severity of PD |
| [107] | characterize metabolite and lipoprotein profiles of newly diagnosed de novo drug-naive PD patients | NMR | serum | - | 329 subjects including de novo drug-naive PD patients, PD patients with advanced disease status, and healthy controls | metabolic differences between newly diagnosed de novo drug-naive PD patients and healthy controls, which were more pronounced in male patients (particularly acetone and cholesterol); metabolic differences between de novo drug-naïve PD patients and advanced PD patients; metabolic differences between advanced PD patients and healthy controls |
| [108] | clinical relevance of microbiome and metabolome alterations in PD | NMR and LC-MS | feces | 16S-sequencing of fecal microbiota | 104 PD patients, 96 control subjects | significant difference in PD gut microbiome and feces metabolome composition; greatest effect size for NMR-based metabolome; SCFAs, lipids, TMAO, ubiquinone and salicylate concentrations vary in PD patients; low SCFA levels |
| | | | | | | correlate with poorer cognition and low BMI; low butyrate levels correlate with worse postural instability-gait disorder scores |
| [109] | integration of longitudinal metabolomics data with constraint-based modeling of gut microbial communities | LC-MS | EDTA plasma | 16S-sequencing of fecal microbiota | 30 PD patients, 30 control subjects | combined omics-methods suggest correlation between sulfur co-metabolism and PD severity; dopaminergic medication affects lipidome; levels of taurine conjugated bile acids correlate with severity of motor symptoms; *A. muciniphila* and *B. wadsworthia* are predicted to alter sulfur metabolism |
| [110] | alterations in gut microbiota might be accompanied by altered concentrations of amino acids, leading to PD | LC-MS, GC-MS | feces | 16S-Sequencing of fecal microbiota | PD patients and healthy controls | greater abundance of *Alistipes*, *Rikenellaceae_RC9_gut_group*, *Bifidobacterium*, *Parabacteroides*, while *Faecalibacterium* was decreased in PD feces specimens; fecal BCAAs and aromatic amino acids concentrations were significantly reduced in PD patients compared to controls |
| [111] | finding a cause-effect relationship between intestinal dysbiosis and PD | GC-MS | feces | 16S-sequencing of fecal microbiota | 64 PD patients, 51 control subjects | alteration of fecal metabolome regarding lipids, amino acids, vitamins, cadaverine, ethanolamine and hydroxy propionic acid; severe metabolomic alterations correlate with abundance of bacteria from the *Lachnospiraceae* family |
| [112] | identification of early biomarkers for PD | FT-ICR-MS | CSF | - | 31 patients, 95 control subjects | 243 metabolites were found to be affected in PD; 15 metabolites are predicted to be the main biological contributors; network analysis showed connection to |
| | | | | | | Krebs-Cycle, possibly displaying mitochondrial dysfunction |
| [113] | integrative metabolic modeling to identify roles of gut microbiota in host metabolism contributing to PD pathophysiology | LC-MS | serum | - | 31 early-stage L-DOPA-naive PD male individuals, 28 matched controls | functional analysis reveals increased microbial capability to degrade mucin and host glycans in PD; personalized community-level metabolic modeling reveals microbial contribution to folate deficiency and hyperhomocysteinemia observed in patients with PD |
| [87] | untargeted metabolomics approach to investigate metabolic changes associated with PD | LC-MS | plasma | - | 223 PD, 169 healthy controls, 68 neurological disease controls | significant reductions of fatty acids and caffeine metabolites, elevation of bile acids; metabolite PD panel with 4 biomarker candidates: FFA10:0, FFA12:0, indolelactic acid and phenylacetyl-glutamine |
| [59] | investigating sebum as potential diagnostic tool for PD; identify PD biomarkers in sebum | LC-MS | sebum | - | 80 drug-naïve PD, 138 medicated PD, 56 healthy controls | 10 metabolites present in samples of drug-naïve and treated PD patients associated with carnitine pathway and sphingolipid metabolism pathway |
| [52] | compare metabolomic profiles of whole blood obtained from treated PD patients, de-novo PD patients and controls, and study perturbations correlated with disease duration, disease stage | GC-MS | blood | - | 16 de-novo PD, 84 treated PD, 42 healthy controls | most prominent differences in butanoic acid and glutamic acid |
| | and motor impairment | | | | | |
| [114] | identify distinct volatiles-associated signature of PD | GC-MS | sebum | - | 43 PD, 21 healthy controls | altered levels of perillic aldehyde, hippuric acid, eicosane, and octadecanal in PD specimens |
| [93] | characterization of metabolic patterns in PD plasma specimens | LC-MS | plasma | - | 28 PD, 18 healthy controls | 17 significantly altered metabolites associated with glycerol phospholipid metabolism, carnitine metabolism, bile acid biosynthesis and tyrosine biosynthesis |
| [53] | identify candidate metabolic biomarker(s) and pathomechanistic pathway(s) of PD | LC-MS | plasma | - | discovery cohort including 82 PD, 82 healthy controls; validation cohort including 118 PD, 22 Huntington's Disease, 47 healthy controls | dopamine and putrescine/ornithine ratio upregulated in PD, octadecadienylcarnitine C18:2, asymmetric dimethylarginine, tryptophan, and kynurenine downregulated in PD |
| [89] | urinary metabolic profiling of idiopathic PD patients at three stages and normal control subjects | GC-MS, LC-MS | urine | - | 92 PD, 65 healthy controls | 18 differential metabolites associated with BCAA metabolism and steroid hormone biosynthesis |
| [69] | identify associations between intestinal microbiome, intestinal digestive function and influence of systemic microbial metabolites on PD | LC-MS | feces, serum | - | 197 PD, 103 healthy controls | different intestinal microbiome composition in PD patients, with increased abundance of *Akkermansia* and *Bifidobacterium* and decreased abundance of *Faecalibacterium* and *Lachnospiraceae*; intestinal microbiome in PD patients had reduced capacity of carbohydrate fermentation and butyrate synthesis and showed increased proteolytic fermentation |
| **(b) Alzheimer's Disease** | | | | | | |
| [97] | investigate role of bile acid composition in AD | LC-MS | serum | - | 1464 subjects (370 cognitively normal, 284 early MCI, 505 late MCI, 305 AD patients) | in AD, cholic acid levels as a primary bile acid are significantly decreased and levels of the secondary bile acid deoxycholic acid are increased; levels of deoxycholic acid conjugated with taurine and glycine are also increased |
| [103] | investigating the metabolic output of gut microbiome dysbiosis in AD | LC-MS | feces | 16S-sequencing of fecal microbiota | 21 patients, 44 control subjects | in AD, 15 gut bacterial genera appear to be altered, 7 of those genera are associated with different series of metabolites; combination of bacterial genera *Faecalibacterium* and *Pseudomonas*, combined with 4 metabolites was able to discriminate between AD patients and controls |
| [96] | identify the relationship between microbiome-associated metabolites and dementia | LC, ion chromatography, GC-MS | feces | classification of fecal bacteria by T-RFLP | 82 control subjects, 25 patients | fecal ammonium and lactic acid were identified as markers for dementia |
| [104] | identify key microbial taxa that participate in the gut-brain axis | CE-FTMS | mouse brain | 16S-sequencing of fecal microbiota | 21 control subjects, 15 patients with MCI, 7 AD patients | *Faecalibacterium prausnitzii* was identified to be participating in the gut-brain axis as its abundance decreased in patients with MCI, correlating with cognitive scores; oral treatment of GMO mice with Aβ-induced cognitive impairment with *F. prausnitzii* improved cognitive impairment and altered metabolic profile in brain tissue specimens |
| [99] | connecting bile acid profiles with standard biomarkers of AD progression | LC-MS | serum | imaging of brain atrophy with MRI, assessment of β-amyloid and tau deposits with PET | 305 control subjects, 98 subjective memory complaint patients, 284 early MCI patients, 505 late MCI patients, 305 AD patients | different bile acid profiles associated with Aβ1-42 in CSF and with p-Tau181 in CSF |
| [98] | metabolomic profiling of bile acids in serum and brain of AD patients | LC-MS | serum and brain tissue | metabolomic analysis of serum and brain samples were also performed in mice | 10 AD patients, 10 healthy subjects | serum levels of cholic acid in AD patients decreased; concentration of taurocholic acid reduced in brain tissue |
| [95] | identification of novel biomarkers for improved risk prediction in AD | LC-MS | serum and brain tissue (post mortem) | follow-up analysis of serum metabolome after 4.5 years | serum: 97 patients with MCI, 433 healthy subjects; brain (post mortem): 28 AD patients, 32 patients | peripheral and systemic metabolome appears to have minor overlaps; three serum acetylcarnitines identified as negative predictors for incident AD and cognitive decline; another 13 metabolites were found as predictors for longitudinal change in cognition |
| | | | | | with MCI, 52 healthy subjects | |

## Metabolic modeling of the gut-brain-axis

A major challenge for microbiome-based approaches especially in neurodegeneration is to deduce molecular mechanisms through which the microbiome could drive disease processes from associations between microbiome composition and disease phenotypes. This is due to the immense complexity of different microbiomes often comprising hundreds to thousands of species with a genetic markup about 150 times larger than that of the host [115,116]. It is further complicated by the large number of factors influencing microbiome composition which make it often impossible to distinguish whether changes in microbiome composition are caused by a disease or are causally involved in its pathogenesis [117,118]. One way to alleviate this problem is the utilization of mechanistic modeling approaches that allow to translate changes in microbiome composition to the potential change in the underlying molecular function of the microbiome [119,120]. One particularly important approach in this regard are constraint-based metabolic modeling approaches that represent individual bacterial taxa as well as the host by their respective metabolic networks [119,120]. Taking into account the nutritional environment of the microbial community, these approaches then use the metabolic networks of the individual species together with an optimization approach to predict metabolic activities in individual species, metabolic exchanges between species and metabolic exchanges with the host. Importantly, these approaches can incorporate compositional information and other types of molecular data such as transcriptomic, proteomic, or metabolomic data, to provide hypotheses about the functional consequences of observed differences in microbiome composition [121–123].

Major approaches that are employed in this context are community flux balance analysis [124], individual-based modeling of microbiome metabolism [125], and whole-body modeling [126]. Community flux balance analysis combines the metabolic networks of individual species into a common compartment and optimizes the total amount of bacterial biomass produced by the entire community [124]. This method assumes that bacterial species are using their metabolic networks such that the entire community produces the highest amount of bacterial biomass possible and hence assumes some intrinsic cooperation in the organization of metabolic fluxes between species. In contrast, individual-based modeling approaches such as BacArena that are also able to account for temporal dynamics, optimize the metabolic networks of bacterial species individually [125]. As a consequence, metabolic interactions in BacArena mostly arise from one bacterial species excreting a product that it does not metabolize further which is taken up by another species that has that capability. In contrast, whole-body modeling aims to build integrated metabolic networks of the host and the metabolic networks of microbial species [126]. These models allow tracing metabolic pathways connecting host and microbiota and thereby are able to propose molecular metabolic pathways through which the microbiome could influence disease processes in the host. Hence, whole-body modeling is able to explicitly model also metabolic interactions along the gastrointestinal-brain-axis.

An additional important feature of these modeling approaches is that they enable the prediction of the outcome of perturbations. Hence they are able to predict specific interventions such as supplementation of nutrients or probiotics that counteract disease-associated microbiome functions and therefore could be an essential component in the rational design of microbiome-based therapies counteracting neurodegeneration.

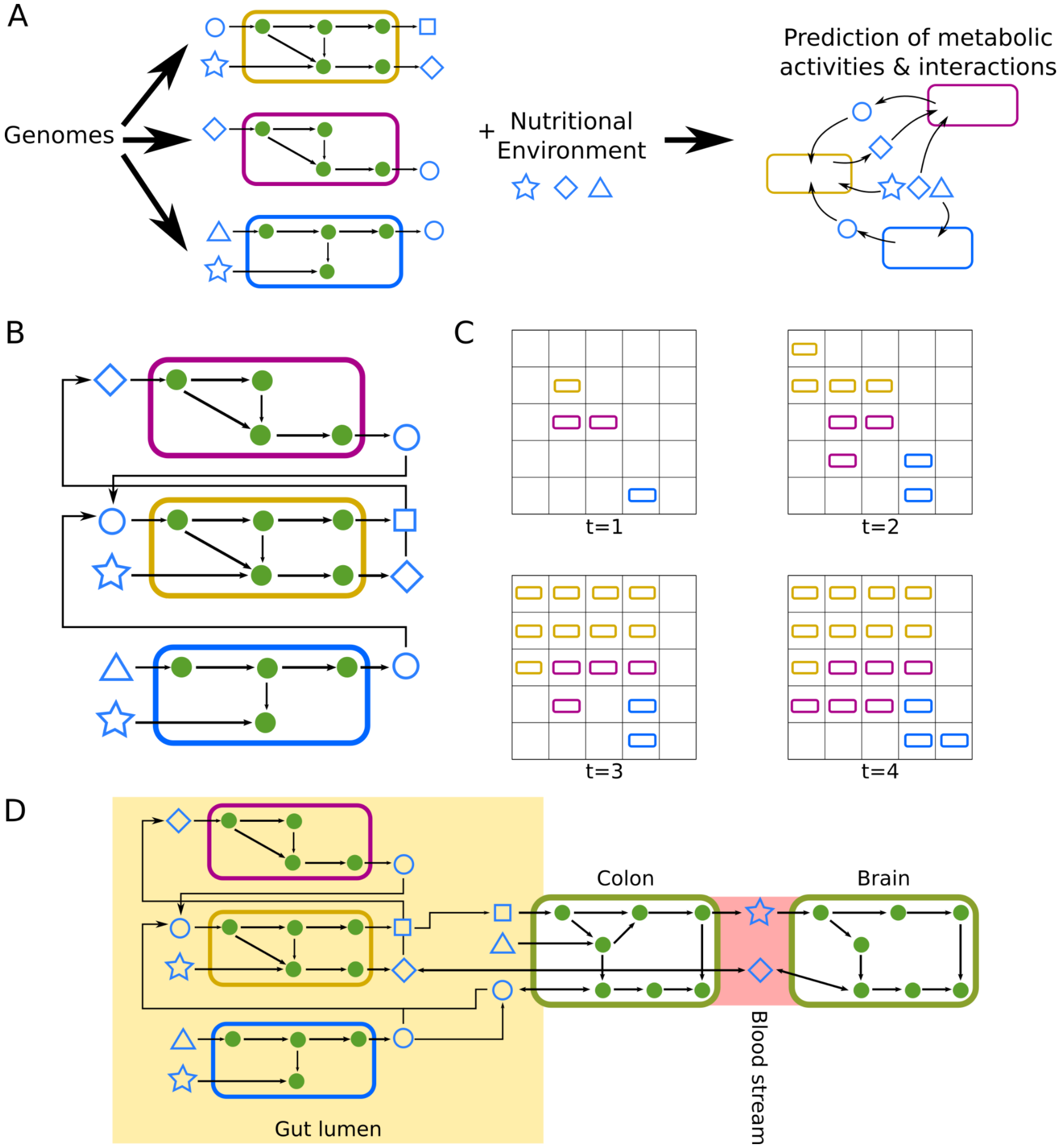

**Fig. 2** Microbial community modeling approaches. Circles correspond to metabolites, arrows to reactions. Shapes indicate exchanged metabolites. **A** General modeling approach. Genomes of bacterial species are translated into their corresponding metabolic network models. Additionally, information on the nutritional environment of the community is added (e.g. reported dietary uptake of a participant). Subsequently, metabolic activities in individual bacterial species and metabolic exchanges between them can be predicted. **B** Community flux balance analysis. Bacterial metabolic networks are combined into a community level metabolic network and it is assumed that bacteria optimize their respective metabolic networks for most efficient community growth. **C** Individual-based modeling of microbial communities. Individual bacterial metabolic networks are simulated in a grid-like environment over time. Metabolic interactions occur as part of the secretion/consumption of metabolites by individual bacteria and diffusion of metabolites between grid cells. **D** Whole-body modeling. Metabolic networks of individual bacteria are joined with metabolic networks representing

individual host tissues. Metabolic exchanges between bacteria and colon occur via a luminal compartment, metabolic exchanges between host tissues are mediated by the blood stream.

### *Microbial community modeling yields insights into neurodegenerative disease-associated changes in microbiome metabolic activity*

Constraint-based microbial community modeling approaches have already seen an application in a large number of different disease contexts, particularly IBD [127,128], type 2 diabetes [129] and PD [113,130]. In the context of IBD, community flux balance analysis was used to assess disease-associated changes in predicted metabolic activities of the microbiome and propose specific metabolic interventions that could counteract these changes [127]. In this study, particular changes in microbial sulfur metabolism were observed which is well in line with metabolomic observations [131]. Moreover, in another study, profound differences in ecological interactions within the microbiome that were predictive of anti-inflammatory therapy success were observed [128]. An interesting link to potential microbiome-based therapeutic approaches was drawn in a recent study investigating the contribution of the microbiome to the therapeutic effect of the anti-diabetic drug metformin [129]. It was found that in the roundworm *Caenorhabditis elegans*, the effect of metformin was mediated by bacterial production of the potential neurotransmitter agmatine. Using microbial community modeling on the gut microbiome of type 2 diabetic patients demonstrated also an increased capacity to produce agmatine in humans taking metformin. Interestingly, agmatine has previously been shown to have a neuroprotective effect in PD [132] as well as in AD [133]. Microbial production of agmatine upon metformin exposure could hence contribute to observed beneficial effects of metformin in AD [134] and PD [135]. In the context of PD, microbial community modeling was used to assess changes in gut microbiome metabolic capacity tied to disease severity [135]. It was found that sulfur-containing compounds such as cysteine-glycine and methionine showed an association with PD. Similarly, another study identified differences in metabolic capacities of individual microbial species that showed an association with PD [113] which were reflected in serum metabolomics data. Interestingly, in the same study, changes in microbial sulfur metabolism were observed.

## The microbiome as therapeutic target in neurodegenerative diseases

The emerging role of the microbiome as a potential driver of neurodegenerative diseases opens up new possibilities for targeted, causal therapies. In this context, there are two fundamentally different approaches to target the microbiome. The first approach are changes in lifestyle that would largely constitute a therapeutic strategy without relevant adverse side-effects or safety concerns, but with high demands regarding personal initiative and adherence. The second therapeutic strategy entails the direct modulation of microbiome composition either through targeted modulation of the abundance of microbial species of interest or a complete replacement of the microbiome through fecal transplants.

### *Changes in Lifestyle: Diet and Exercise*

As discussed, diet is one of the strongest factors influencing microbiome composition [136]. For neurodegenerative diseases, a variety of animal and observational studies have indicated beneficial effects of different forms of diet and nutritional habits. Of particular interest is the Mediterranean diet, which is associated with a decreased risk for PD and AD [137]. Parts of the neuroprotective or anti-inflammatory effects of the Mediterranean diet could be mediated by the microbiome [138]. One important aspect of this diet is the increased intake of fibers, which forms a direct link to microbially produced SCFAs [11]. Apart from effects on the intestinal and endocrinological system, SCFAs have been associated with positive effects on immunologic functions, including modulation of microglia and T-cell function in the ENS and CNS [139]. However, their specific role in neurodegenerative diseases remains somewhat convoluted [6]. Another important constituent of the Mediterranean diet are secondary plant compounds such as polyphenoles. 90-95 % of total polyphenol intake may accumulate in the large intestine, where they become available for fermentation by the gut microbiota. Polyphenols and their degradation products (e.g. hydroxybenzoic acids) have been reported to inhibit the formation of alpha-synuclein misfolded aggregates, reduce mitochondrial dysfunction-induced oxidative stress, and inflammatory responses [140,141]. In contrast, the *Western diet,* including highly-processed, high-fat and high-sugar foods, has been associated with pro-inflammatory properties, which have been linked to AD pathology and an increased risk for PD [142] with part of these effects seemingly mediated by the microbiome [143]. The ketogenic diet has also been considered for its potential health benefits in neurological diseases, including PD, for which rodent models indicate that effects might be mediated by the gut microbiome [144]. Results from cohort studies and meta-analyses that focus on single food items like dairy products and alcohol or single nutrients like calcium, antioxidants, B-vitamins and n-6 or n-3 polyunsaturated fatty acids have shown inconsistent results (for a review see [145]). Since diet is a multidimensional exposure of components with different health effects, a diet intervention based on an individual's dietary patterns might have more favorable effects if it alters intake of multiple foods that may lead to a combination of many smaller effect sizes [146]. On the other hand, health effects of dietary patterns may depend on genetic risk alleles like the APOE4 genotype as well as on the gut microbiome. Compared to vegans, omnivores produce significantly higher levels of the atherosclerosis-promoting trimethylamine-N-oxide (TMAO) after eating a protein-rich meal because several bacterial taxa that form the TMAO precursor trimethylamine have been reported to be more abundant in omnivores than in vegans [147]. The Nutrition for Dementia Prevention Working Group recently proposed a roadmap for future studies in nutrition and dementia prevention [146]. According to their recommendations, diets should be designed based on multiple neuroprotective dietary or nutrient components that can be applied in interventional trials. In addition, smaller personalized trials should be performed that consider genetics, omics, microbiome, and nutrient exposures and are guided by biomarkers that reflect brain functions.

Another important aspect of lifestyle as a therapeutic option for targeting the microbiome is exercise which was shown to decrease the risk for neurodegeneration, induce neurorestorative and neuroprotective effects and modulate disease progression in animal and human observational studies [148,149]. In this respect, a variety of rodent model studies revealed alterations of the gut microbiome following different forms of exercise, in interaction with but also independently from dietary changes [150,151]. Similar effects have been observed in human studies, with alterations of microbial diversity observed in athletes and following different forms of exercise [152]. However, clinical interventional studies linking exercise-induced changes in the microbiome and direct health benefits in neurodegenerative diseases are still missing. Additionally, especially for PD the interaction between exercise effects on gut motility and changes in the intestinal flora should be further elucidated.

Taken together, evidence from human studies confirming a direct link between lifestyle changes, microbiome alterations and clinical benefits in neurodegenerative diseases are still needed. Moreover, while the potential therapeutic role of SCFAs needs to be clarified, available studies strongly support positive effects of a high-fiber, Mediterranean diet and regular exercise, which should be mechanistically further elucidated to specifically advise them as low-risk therapeutic options for neurodegenerative diseases.

***Prebiotics and Probiotics***

Following accumulating evidence of a prominent role of the GI-brain axis in neurodegenerative diseases, researchers and patients placed high hopes in the use of *prebiotics* (nutrients supporting beneficial microbial strains), *probiotics* (beneficial strains) or their combination referred to as *synbiotics*, to target the microbiome. In this respect the use of prebiotics has a high overlap to positively rated foods from dietary studies. Important prebiotics include fructooligosaccharides (FOS), galactooligosaccharides (GOS), polyunsaturated fatty acids (PUFA) and plant polyphenols. For neurodegenerative diseases, especially the neuroprotective, anti-inflammatory and antioxidative effects of polyphenols and PUFAs have been described [140,153]. So far, a variety of animal and human studies have examined behavioral and neuropsychiatric effects of prebiotics, including effects on anxiety, depression and memory function [154,155]. In AD mouse models, prebiotics such as plant polyphenols and oligosaccharides have been linked to an improvement of cognitive function and modulation of amyloid or tau pathology, in association with microbiome diversity and metabolism [156]. In contrast, very limited studies have been performed for PD, examining mainly the role of prebiotics (and probiotics) on constipation in clinical PD [157]. Concerning probiotics, various animal and few human studies have shown behavioral and neuropsychiatric effects including effects on cognitive function or fatigue [158,159]. Accordingly, first interventional studies in AD and PD have been conducted, showing potential clinical benefits of microbiome alterations [160,161]. In summary, despite great interest in the use of prebiotics and probiotics to target microbiome-associated disease progression in neurodegeneration and first promising results, evidence from clinical studies is not sufficient, yet, for an official medical recommendation to use probiotics or prebiotics in AD or PD.

***Antibiotics***

The other route besides proliferation promotion with pre- and probiotics is the suppression of invasive or overabundant species through antibiotics. The treatment with broad-spectrum antibiotics is considered to have severe microbiome-related side effects such as microbiome dysbiosis [162]. For example, such a treatment has decreased the survival rates of patients with cancer [163] indicating the role of homeostasis for the overall health of a patient. Also, intake of antibiotic medication has been suggested to increase PD risk in healthy individuals [164]. Therefore, caution is warranted with more general antibiotic treatment, while more targeted antibiotic interventions might be more promising. For instance, the overabundance of specific species can trigger the release of inflammatory mediators in AD patients, e.g., *Helicobacter pylori* [165]. Even viral infections like herpes simplex virus (HSV) type 1 have been identified as possible risk factors in AD [166]. Antibiotic targeting of individual species inside the microbiome could cut feedback loops and synergistic effects important for the modulation of the overarching disease. For chronic peptic ulcers, this has already been implemented for a long time by specifically targeting *H. pylori* [167]. Another example is the treatment of *Clostridioides difficile* infections (CDI), where the switch from a broad-spectrum antibiotic to a specifically targeting, microbiome-sparing antibiotic could reduce CDI recurrence levels by 60% compared to the standard clinical therapy [168].

Still, besides interfering with the biochemical pathways of the pathogens, solid-state structural mechanisms could be proposed as well. Specifically antibiotic, but biocompatible particles like zinc oxide micro-tetrapods [169] have already been employed to facilitate an immune response against HSV [170–173]. In this case, instead of a classic pharmaceutical effect through the release of zinc ions, the structural effect of binding virion glycoprotein groups to designed surface oxygen vacancies is used. For HSV the experimental evidence elucidated the mechanism: a capturing of the nanoscopic virion to the microscopic tetrapodal zinc oxide surface was observed. Thus, the tetrapods acted as a virostatic platform. From there, antigen-presenting cells identified the immobilized viruses and thus triggered the immune system via the CD4/CD8 signaling pathway against herpes simplex viruses in a mouse model. As herpes simplex is one risk factor for AD, such a therapy could also reduce the overall AD risk.

This solid-state structural strategy is also applicable for bacteria. Structural differences cause differences in the antibacterial efficacy [170–172]. By using specific binding elements on top of the micro crystals, these can be tailored even more for a microbiome-sparing antibiotic targeting. By chemically altering the surface structure of the tetrapodal zinc oxide particles, the binding specificity towards other proteins for simultaneously anti- and prebiotic purposes could be achieved [170–172]. Such a combined approach may significantly impact the species in the microbiome as well as break negative and enhance positive feedback towards homeostasis.

***Fecal transplants***
The possibility to alter the microbiome via fecal transplants has gained a lot of attention as a potential therapeutic option in neurodegenerative diseases. Fecal transplants are applied evidence-based, recommended for recurrent clostridium difficile infections [174]. Mouse models have revealed effects of FMT on neurobehavior [175]. Moreover, of interest for pathological pathways in AD and PD, alterations of the immune system by FMT have been observed in the mouse model [176]. First small clinical studies and case reports addressed effects of FMT on gastrointestinal symptoms in PD [177]. So far, disease-modifying effects of stool transplants in neurodegenerative diseases have not been reported.

***Medication***
Apart from the well-known effects of antibiotics on the microbiome [178], a multitude of studies showed that other drug classes, including antidiabetics (metformin), psychiatric medication (antidepressants) or proton pump inhibitors influence the gut microbiome [129,179,180]. In turn, effects of patient medication can be compromised by gut microbiota metabolization as shown for L-dopa in PD [181]. However, further studies are needed to determine whether specific medication can be used to yield microbiome-mediated positive effects on neurodegenerative diseases.

Taken together the microbiome is an interesting therapeutic target, especially considering the possibility of easily applicable low-risk interventional therapies. However, it has to be taken into account that most of the microbiome studies in AD and PD have been conducted in the clinical phase of the diseases, with the goal of disease-modification. Causal treatment, however, should directly target underlying pathomechanisms occuring in the prodromal/preclinical phases of neurodegeneration, which, in terms of microbiome alterations, have yet to be elucidated. So far, the evidence for a medical use of probiotics/prebiotics or FMT is not sufficient, especially considering rare but possible side-effects or adverse reactions of these treatments. However, considering the wide-ranging beneficial effects of a healthy lifestyle, recommendations on diet and exercise, potentially influencing microbiome-driven pathology in PD and AD, should be further explored.

## Summary and future research perspectives

Both PD and AD, the two most common neurodegenerative diseases, have been associated with substantial changes in the microbiome and metabolome composition in comparison to healthy individuals. Nevertheless, the prognostic value of these changes for early disease diagnosis and prognosis, most particularly in the prodromal disease phases, still needs to be determined in large-scale, prospective multiomics studies. On the other hand, both the microbiome and the metabolome might offer novel therapeutic targets for effective disease treatment. This, however, requires a deep, systematic understanding of the underlying disease pathomechanisms to improve patient outcomes and minimize side-effects. A systems-wide understanding of the GI-brain axis covering not only individual organs and/or biocompartments, but comprising the whole organism would potentially facilitate these insights. Systems Medicine and Systems Biological modeling approaches of host-microbiome-interactions represent key tools in achieving such a systematic understanding. On the sampling side, however, both microbiome and metabolome analyses in PD and AD have been limited to individual organs and mainly standard biospecimens like feces and blood. The additional multiomics investigation of other biospecimens such as saliva and CSF might, in particular, offer novel insights into the link between the oral microbiome, the GI-brain axis, and neurodegeneration. Many options to actively change the microbiome and metabolome composition are already available, most importantly dietary interventions and lifestyle changes. But their downstream effects on neurodegeneration still need to be explored. With respect to the long prodromal phases of PD and AD, a major focus should be set on affordable and low-threshold interventions.

Finally, systematic comparisons of microbiome and metabolome changes across different neurodegenerative diseases both cross-sectionally and longitudinally are still lacking. An extension of these systematic comparisons to other, at least partially neurodegenerative diseases, such as multiple sclerosis might unveil disease-overarching pathomechanisms of neurodegeneration. In fact, profound microbiome and metabolome changes have already been reported in multiple sclerosis [182,183]. Thus, deeply phenotyped multi-cohort studies as well as clinical trials including several distinct neurodegenerative diseases such as PD, AD, and multiple sclerosis might pave the way to a deeper mechanistic understanding of neurodegeneration and uncover novel therapeutic strategies to fight this global pandemic.

## Acknowledgement

HUZ is supported by the Federal Ministry of Education and Research (Bundesministerium für Bildung und Forschung; BMBF) within the framework of the e:Med research and funding concept (grant 01ZX1912A). CK acknowledges support by the German Research Foundation (CRC 1182 Metaorganisms and KA 3541/18-1).

## References

1. Kivipelto M, Mangialasche F, Ngandu T. Lifestyle interventions to prevent cognitive impairment, dementia and Alzheimer disease. Nat Rev Neurol. 2018;14: 653–666.
2. Ascherio A, Schwarzschild MA. The epidemiology of Parkinson's disease: risk factors and prevention. Lancet Neurol. 2016;15: 1257–1272.
3. Dorsey ER, Sherer T, Okun MS, Bloem BR. The Emerging Evidence of the Parkinson Pandemic. J Parkinsons Dis. 2018;8: S3–S8.
4. GBD 2019 Dementia Forecasting Collaborators. Estimation of the global prevalence of dementia in 2019 and forecasted prevalence in 2050: an analysis for the Global Burden of Disease Study 2019. Lancet Public Health. 2022;7: e105–e125.
5. Carabotti M, Scirocco A, Maselli MA, Severi C. The gut-brain axis: interactions between enteric microbiota, central and enteric nervous systems. Ann Gastroenterol Hepatol . 2015;28: 203–209.
6. Rosario D, Boren J, Uhlen M, Proctor G, Aarsland D, Mardinoglu A, et al. Systems Biology Approaches to Understand the Host-Microbiome Interactions in Neurodegenerative Diseases. Front Neurosci. 2020;14: 716.
7. Fan Y, Pedersen O. Gut microbiota in human metabolic health and disease. Nat Rev Microbiol. 2021;19: 55–71.
8. Morais LH, Schreiber HL 4th, Mazmanian SK. The gut microbiota-brain axis in behaviour and brain disorders. Nat Rev Microbiol. 2021;19: 241–255.
9. Hoyles L, Snelling T, Umlai U-K, Nicholson JK, Carding SR, Glen RC, et al. Microbiome-host systems interactions: protective effects of propionate upon the blood-brain barrier. Microbiome. 2018;6: 55.
10. Topping DL, Clifton PM. Short-chain fatty acids and human colonic function: roles of resistant starch and nonstarch polysaccharides. Physiol Rev. 2001;81: 1031–1064.
11. Koh A, De Vadder F, Kovatcheva-Datchary P, Bäckhed F. From Dietary Fiber to Host Physiology: Short-Chain Fatty Acids as Key Bacterial Metabolites. Cell. 2016;165: 1332–1345.
12. Roediger WE. The starved colon--diminished mucosal nutrition, diminished absorption, and colitis. Dis Colon Rectum. 1990;33: 858–862.
13. Lozupone CA, Knight R. Global patterns in bacterial diversity. Proc Natl Acad Sci U S A. 2007;104: 11436–11440.
14. Tyson GW, Chapman J, Hugenholtz P, Allen EE, Ram RJ, Richardson PM, et al. Community structure and metabolism through reconstruction of microbial genomes from the environment. Nature. 2004;428: 37–43.
15. Zhernakova A, Kurilshikov A, Bonder MJ, Tigchelaar EF, Schirmer M, Vatanen T, et al. Population-based metagenomics analysis reveals markers for gut microbiome composition and diversity. Science. 2016;352: 565–569.
16. Pasolli E, Asnicar F, Manara S, Zolfo M, Karcher N, Armanini F, et al. Extensive Unexplored Human Microbiome Diversity Revealed by Over 150,000 Genomes from Metagenomes Spanning Age, Geography, and Lifestyle. Cell. 2019;176: 649–662.e20.
17. Olm MR, Crits-Christoph A, Bouma-Gregson K, Firek BA, Morowitz MJ, Banfield JF. inStrain profiles population microdiversity from metagenomic data and sensitively detects shared microbial strains. Nat Biotechnol. 2021;39: 727–736.
18. Poyet M, Groussin M, Gibbons SM, Avila-Pacheco J, Jiang X, Kearney SM, et al. A library of human gut bacterial isolates paired with longitudinal multiomics data enables mechanistic microbiome research. Nat Med. 2019;25: 1442–1452.
19. Almeida A, Nayfach S, Boland M, Strozzi F, Beracochea M, Shi ZJ, et al. A unified catalog of 204,938 reference genomes from the human gut microbiome. Nat Biotechnol. 2021;39: 105–114.
20. Gligorijević V, Renfrew PD, Kosciolek T, Leman JK, Berenberg D, Vatanen T, et al. Structure-based protein function prediction using graph convolutional networks. Nat Commun. 2021;12: 3168.
21. Gacesa R, Kurilshikov A, Vich Vila A, Sinha T, Klaassen MAY, Bolte LA, et al. Environmental factors shaping the gut microbiome in a Dutch population. Nature. 2022;604: 732–739.
22. Morton JT, Aksenov AA, Nothias LF, Foulds JR, Quinn RA, Badri MH, et al. Learning representations of microbe-metabolite interactions. Nat Methods. 2019;16: 1306–1314.
23. Duvallet C, Gibbons SM, Gurry T, Irizarry RA, Alm EJ. Meta-analysis of gut microbiome studies identifies disease-specific and shared responses. Nat Commun. 2017;8: 1784.
24. McDonald D, Hyde E, Debelius JW, Morton JT, Gonzalez A, Ackermann G, et al. American Gut: an Open Platform for Citizen Science Microbiome Research. mSystems. 2018;3. doi:10.1128/mSystems.00031-18
25. He Y, Wu W, Zheng H-M, Li P, McDonald D, Sheng H-F, et al. Regional variation limits applications of healthy gut microbiome reference ranges and disease models. Nat Med. 2018;24: 1532–1535.
26. Groussin M, Poyet M, Sistiaga A, Kearney SM, Moniz K, Noel M, et al. Elevated rates of horizontal gene transfer in the industrialized human microbiome. Cell. 2021;184: 2053–2067.e18.
27. Costea PI, Zeller G, Sunagawa S, Pelletier E, Alberti A, Levenez F, et al. Towards standards for human fecal sample processing in metagenomic studies. Nat Biotechnol. 2017;35: 1069–1076.
28. Sinha R, The Microbiome Quality Control Project Consortium, Abu-Ali G, Vogtmann E, Fodor AA, Ren B, et al. Assessment of variation in microbial community amplicon sequencing by the Microbiome Quality Control (MBQC) project consortium. Nature Biotechnology. 2017. pp. 1077–1086. doi:10.1038/nbt.3981
29. Thorsen J, Brejnrod A, Mortensen M, Rasmussen MA, Stokholm J, Al-Soud WA, et al. Large-scale benchmarking reveals false discoveries and count transformation sensitivity in 16S rRNA gene amplicon data analysis methods used in microbiome studies. Microbiome. 2016;4: 62.
30. Mirzayi C, Renson A, Genomic Standards Consortium, Massive Analysis and Quality Control Society, Zohra

F, Elsafoury S, et al. Reporting guidelines for human microbiome research: the STORMS checklist. Nat Med. 2021;27: 1885–1892.
31. Seo D-O, Holtzman DM. Gut Microbiota: From the Forgotten Organ to a Potential Key Player in the Pathology of Alzheimer's Disease. J Gerontol A Biol Sci Med Sci. 2020;75: 1232–1241.
32. Friedland RP, Chapman MR. The role of microbial amyloid in neurodegeneration. PLoS Pathog. 2017;13: e1006654.
33. Glassner KL, Abraham BP, Quigley EMM. The microbiome and inflammatory bowel disease. J Allergy Clin Immunol. 2020;145: 16–27.
34. Romano S, Savva GM, Bedarf JR, Charles IG, Hildebrand F, Narbad A. Meta-analysis of the Parkinson's disease gut microbiome suggests alterations linked to intestinal inflammation. NPJ Parkinsons Dis. 2021;7: 27.
35. Jiang C, Li G, Huang P, Liu Z, Zhao B. The Gut Microbiota and Alzheimer's Disease. J Alzheimers Dis. 2017;58: 1–15.
36. Smith P, Willemsen D, Popkes M, Metge F, Gandiwa E, Reichard M, et al. Regulation of life span by the gut microbiota in the short-lived African turquoise killifish. Elife. 2017;6. doi:10.7554/eLife.27014
37. Walter J, Armet AM, Finlay BB, Shanahan F. Establishing or Exaggerating Causality for the Gut Microbiome: Lessons from Human Microbiota-Associated Rodents. Cell. 2020;180: 221–232.
38. Zacharias HU, Altenbuchinger M, Gronwald W. Statistical Analysis of NMR Metabolic Fingerprints: Established Methods and Recent Advances. Metabolites. 2018;8. doi:10.3390/metabo8030047
39. Bauermeister A, Mannochio-Russo H, Costa-Lotufo LV, Jarmusch AK, Dorrestein PC. Mass spectrometry-based metabolomics in microbiome investigations. Nat Rev Microbiol. 2022;20: 143–160.
40. Keppler EAH, Jenkins CL, Davis TJ, Bean HD. Advances in the application of comprehensive two-dimensional gas chromatography in metabolomics. Trends Analyt Chem. 2018;109: 275–286.
41. Papadimitropoulos M-EP, Vasilopoulou CG, Maga-Nteve C, Klapa MI. Untargeted GC-MS Metabolomics. Methods Mol Biol. 2018;1738: 133–147.
42. Herderich M, Richling E, Roscher R, Schneider C, Schwab W, Humpf H-U, et al. Application of atmospheric pressure ionization HPLC-MS-MS for the analysis of natural products. Chromatographia. 1997. pp. 127–132. doi:10.1007/bf02505549
43. Want EJ, Nordström A, Morita H, Siuzdak G. From exogenous to endogenous: the inevitable imprint of mass spectrometry in metabolomics. J Proteome Res. 2007;6: 459–468.
44. Pasikanti KK, Ho PC, Chan ECY. Gas chromatography/mass spectrometry in metabolic profiling of biological fluids. J Chromatogr B Analyt Technol Biomed Life Sci. 2008;871: 202–211.
45. Jiménez B, Holmes E, Heude C, Tolson RF, Harvey N, Lodge SL, et al. Quantitative Lipoprotein Subclass and Low Molecular Weight Metabolite Analysis in Human Serum and Plasma by H NMR Spectroscopy in a Multilaboratory Trial. Anal Chem. 2018;90: 11962–11971.
46. Lei Z, Huhman DV, Sumner LW. Mass spectrometry strategies in metabolomics. J Biol Chem. 2011;286: 25435–25442.
47. Erve JCL, Demaio W, Talaat RE. Rapid metabolite identification with sub parts-per-million mass accuracy from biological matrices by direct infusion nanoelectrospray ionization after clean-up on a ZipTip and LTQ/Orbitrap mass spectrometry. Rapid Commun Mass Spectrom. 2008;22: 3015–3026.
48. Beckonert O, Keun HC, Ebbels TMD, Bundy J, Holmes E, Lindon JC, et al. Metabolic profiling, metabolomic and metabonomic procedures for NMR spectroscopy of urine, plasma, serum and tissue extracts. Nat Protoc. 2007;2: 2692–2703.
49. Zacharias H, Hochrein J, Klein M, Samol C, Oefner P, Gronwald W. Current experimental, bioinformatic and statistical methods used in NMR based metabolomics. Curr Metabolomics. 2013;1: 253–268.
50. Schultheiss UT, Kosch R, Kotsis F, Altenbuchinger M, Zacharias HU. Chronic Kidney Disease Cohort Studies: A Guide to Metabolome Analyses. Metabolites. 2021;11. doi:10.3390/metabo11070460
51. Trezzi J-P, Galozzi S, Jaeger C, Barkovits K, Brockmann K, Maetzler W, et al. Distinct metabolomic signature in cerebrospinal fluid in early parkinson's disease. Mov Disord. 2017;32: 1401–1408.
52. Troisi J, Landolfi A, Vitale C, Longo K, Cozzolino A, Squillante M, et al. A metabolomic signature of treated and drug-naïve patients with Parkinson's disease: a pilot study. Metabolomics. 2019;15: 90.
53. Chang K-H, Cheng M-L, Tang H-Y, Huang C-Y, Wu Y-R, Chen C-M. Alternations of Metabolic Profile and Kynurenine Metabolism in the Plasma of Parkinson's Disease. Mol Neurobiol. 2018;55: 6319–6328.
54. Michell AW, Mosedale D, Grainger DJ, Barker RA. Metabolomic analysis of urine and serum in Parkinson's disease. Metabolomics. 2008;4: 191–201.
55. Li X, Fan X, Yang H, Liu Y. Review of Metabolomics-Based Biomarker Research for Parkinson's Disease. Mol Neurobiol. 2022;59: 1041–1057.
56. Unger MM, Spiegel J, Dillmann K-U, Grundmann D, Philippeit H, Bürmann J, et al. Short chain fatty acids and gut microbiota differ between patients with Parkinson's disease and age-matched controls. Parkinsonism Relat Disord. 2016;32: 66–72.
57. Shao Y, Le W. Recent advances and perspectives of metabolomics-based investigations in Parkinson's disease. Mol Neurodegener. 2019;14: 3.
58. Kumari S, Goyal V, Kumaran SS, Dwivedi SN, Srivastava A, Jagannathan NR. Quantitative metabolomics of saliva using proton NMR spectroscopy in patients with Parkinson's disease and healthy controls. Neurol Sci. 2020;41: 1201–1210.
59. Sinclair E, Trivedi DK, Sarkar D, Walton-Doyle C, Milne J, Kunath T, et al. Metabolomics of sebum reveals lipid dysregulation in Parkinson's disease. Nat Commun. 2021;12: 1592.

60. Bouatra S, Aziat F, Mandal R, Guo AC, Wilson MR, Knox C, et al. The human urine metabolome. PLoS One. 2013;8: e73076.
61. Bogdanov M, Matson WR, Wang L, Matson T, Saunders-Pullman R, Bressman SS, et al. Metabolomic profiling to develop blood biomarkers for Parkinson's disease. Brain. 2008;131: 389–396.
62. Zhao L, Ni Y, Su M, Li H, Dong F, Chen W, et al. High Throughput and Quantitative Measurement of Microbial Metabolome by Gas Chromatography/Mass Spectrometry Using Automated Alkyl Chloroformate Derivatization. Anal Chem. 2017;89: 5565–5577.
63. Plassais J, Gbikpi-Benissan G, Figarol M, Scheperjans F, Gorochov G, Derkinderen P, et al. Gut microbiome alpha-diversity is not a marker of Parkinson's disease and multiple sclerosis. Brain Commun. 2021;3: fcab113.
64. Toh TS, Chong CW, Lim S-Y, Bowman J, Cirstea M, Lin C-H, et al. Gut microbiome in Parkinson's disease: New insights from meta-analysis. Parkinsonism Relat Disord. 2022;94: 1–9.
65. Heintz-Buschart A, Pandey U, Wicke T, Sixel-Döring F, Janzen A, Sittig-Wiegand E, et al. The nasal and gut microbiome in Parkinson's disease and idiopathic rapid eye movement sleep behavior disorder. Mov Disord. 2018;33: 88–98.
66. Heinzel S, Aho VTE, Suenkel U, von Thaler A-K, Schulte C, Deuschle C, et al. Gut Microbiome Signatures of Risk and Prodromal Markers of Parkinson Disease. Ann Neurol. 2021;90: E1–E12.
67. Qian Y, Yang X, Xu S, Huang P, Li B, Du J, et al. Gut metagenomics-derived genes as potential biomarkers of Parkinson's disease. Brain. 2020;143: 2474–2489.
68. Mao L, Zhang Y, Tian J, Sang M, Zhang G, Zhou Y, et al. Cross-Sectional Study on the Gut Microbiome of Parkinson's Disease Patients in Central China. Front Microbiol. 2021;12: 728479.
69. Cirstea MS, Yu AC, Golz E, Sundvick K, Kliger D, Radisavljevic N, et al. Microbiota Composition and Metabolism Are Associated With Gut Function in Parkinson's Disease. Mov Disord. 2020;35: 1208–1217.
70. Aho VTE, Houser MC, Pereira PAB, Chang J, Rudi K, Paulin L, et al. Relationships of gut microbiota, short-chain fatty acids, inflammation, and the gut barrier in Parkinson's disease. Mol Neurodegener. 2021;16: 6.
71. Shin C, Lim Y, Lim H, Ahn T-B. Plasma Short-Chain Fatty Acids in Patients With Parkinson's Disease. Mov Disord. 2020;35: 1021–1027.
72. He X, Qian Y, Xu S, Zhang Y, Mo C, Guo W, et al. Plasma Short-Chain Fatty Acids Differences in Multiple System Atrophy from Parkinson's Disease. J Parkinsons Dis. 2021;11: 1167–1176.
73. Chen S-J, Chen C-C, Liao H-Y, Lin Y-T, Wu Y-W, Liou J-M, et al. Association of Fecal and Plasma Levels of Short-Chain Fatty Acids With Gut Microbiota and Clinical Severity in Patients With Parkinson Disease. Neurology. 2022;98: e848–e858.
74. Bairamian D, Sha S, Rolhion N, Sokol H, Dorothée G, Lemere CA, et al. Microbiota in neuroinflammation and synaptic dysfunction: a focus on Alzheimer's disease. Mol Neurodegener. 2022;17: 1–23.
75. Mariat D, Firmesse O, Levenez F, Guimarăes VD, Sokol H, Doré J, et al. The Firmicutes/Bacteroidetes ratio of the human microbiota changes with age. BMC Microbiol. 2009;9: 1–6.
76. Vogt NM, Kerby RL, Dill-McFarland KA, Harding SJ, Merluzzi AP, Johnson SC, et al. Gut microbiome alterations in Alzheimer's disease. Sci Rep. 2017;7: 1–11.
77. Shukla PK, Delotterie DF, Xiao J, Pierre JF, Rao R, McDonald MP, et al. Alterations in the Gut-Microbial-Inflammasome-Brain Axis in a Mouse Model of Alzheimer's Disease. Cells. 2021;10: 779.
78. Sun B-L, Li W-W, Wang J, Xu Y-L, Sun H-L, Tian D-Y, et al. Gut Microbiota Alteration and Its Time Course in a Tauopathy Mouse Model. J Alzheimers Dis. 2019;70: 399–412.
79. Erny D, Hrabě de Angelis AL, Jaitin D, Wieghofer P, Staszewski O, David E, et al. Host microbiota constantly control maturation and function of microglia in the CNS. Nat Neurosci. 2015;18: 965–977.
80. Silva YP, Bernardi A, Frozza RL. The Role of Short-Chain Fatty Acids From Gut Microbiota in Gut-Brain Communication. Front Endocrinol . 2020;0. doi:10.3389/fendo.2020.00025
81. Marizzoni M, Cattaneo A, Mirabelli P, Festari C, Lopizzo N, Nicolosi V, et al. Short-Chain Fatty Acids and Lipopolysaccharide as Mediators Between Gut Dysbiosis and Amyloid Pathology in Alzheimer's Disease. J Alzheimers Dis. 2020;78: 683–697.
82. Cuervo-Zanatta D, Syeda T, Sánchez-Valle V, Irene-Fierro M, Torres-Aguilar P, Torres-Ramos MA, et al. Dietary Fiber Modulates the Release of Gut Bacterial Products Preventing Cognitive Decline in an Alzheimer's Mouse Model. Cell Mol Neurobiol. 2022; 1–24.
83. Shi Y, Holtzman DM. Interplay between innate immunity and Alzheimer disease: APOE and TREM2 in the spotlight. Nat Rev Immunol. 2018;18: 759–772.
84. Tran TTT, Corsini S, Kellingray L, Hegarty C, Le Gall G, Narbad A, et al. genotype influences the gut microbiome structure and function in humans and mice: relevance for Alzheimer's disease pathophysiology. FASEB J. 2019;33: 8221–8231.
85. Friedland RP, McMillan J, Kurlawala Z. What Are the Molecular Mechanisms by Which Functional Bacterial Amyloids Influence Amyloid Beta Deposition and Neuroinflammation in Neurodegenerative Disorders? Int J Mol Sci. 2020;21: 1652.
86. Holmqvist S, Chutna O, Bousset L, Aldrin-Kirk P, Li W, Björklund T, et al. Direct evidence of Parkinson pathology spread from the gastrointestinal tract to the brain in rats. Acta Neuropathol. 2014;128: 805–820.
87. Shao Y, Li T, Liu Z, Wang X, Xu X, Li S, et al. Comprehensive metabolic profiling of Parkinson's disease by liquid chromatography-mass spectrometry. Mol Neurodegener. 2021;16: 4.
88. Hatano T, Saiki S, Okuzumi A, Mohney RP, Hattori N. Identification of novel biomarkers for Parkinson's disease by metabolomic technologies. J Neurol Neurosurg Psychiatry. 2016;87: 295–301.

89. Luan H, Liu L-F, Tang Z, Zhang M, Chua K-K, Song J-X, et al. Comprehensive urinary metabolomic profiling and identification of potential noninvasive marker for idiopathic Parkinson's disease. Sci Rep. 2015;5: 13888.
90. Olsson E, Byberg L, Höijer J, Kilander L, Larsson SC. Milk and Fermented Milk Intake and Parkinson's Disease: Cohort Study. Nutrients. 2020;12. doi:10.3390/nu12092763
91. Jiang W, Ju C, Jiang H, Zhang D. Dairy foods intake and risk of Parkinson's disease: a dose-response meta-analysis of prospective cohort studies. Eur J Epidemiol. 2014;29: 613–619.
92. Tosukhowong P, Boonla C, Dissayabutra T, Kaewwilai L, Muensri S, Chotipanich C, et al. Biochemical and clinical effects of Whey protein supplementation in Parkinson's disease: A pilot study. J Neurol Sci. 2016;367: 162–170.
93. Zhao H, Wang C, Zhao N, Li W, Yang Z, Liu X, et al. Potential biomarkers of Parkinson's disease revealed by plasma metabolic profiling. J Chromatogr B Analyt Technol Biomed Life Sci. 2018;1081-1082: 101–108.
94. Wang Y, Chen S, Tan J, Gao Y, Yan H, Liu Y, et al. Tryptophan in the diet ameliorates motor deficits in a rotenone-induced rat Parkinson's disease model via activating the aromatic hydrocarbon receptor pathway. Brain Behav. 2021;11: e2226.
95. Huo Z, Yu L, Yang J, Zhu Y, Bennett DA, Zhao J. Brain and blood metabolome for Alzheimer's dementia: findings from a targeted metabolomics analysis. Neurobiol Aging. 2020;86: 123–133.
96. Saji N, Murotani K, Hisada T, Kunihiro T, Tsuduki T, Sugimoto T, et al. Relationship between dementia and gut microbiome-associated metabolites: a cross-sectional study in Japan. Sci Rep. 2020;10: 8088.
97. MahmoudianDehkordi S, Arnold M, Nho K, Ahmad S, Jia W, Xie G, et al. Altered bile acid profile associates with cognitive impairment in Alzheimer's disease-An emerging role for gut microbiome. Alzheimers Dement. 2019;15: 76–92.
98. Pan X, Elliott CT, McGuinness B, Passmore P, Kehoe PG, Hölscher C, et al. Metabolomic Profiling of Bile Acids in Clinical and Experimental Samples of Alzheimer's Disease. Metabolites. 2017;7. doi:10.3390/metabo7020028
99. Nho K, Kueider-Paisley A, MahmoudianDehkordi S, Arnold M, Risacher SL, Louie G, et al. Altered bile acid profile in mild cognitive impairment and Alzheimer's disease: Relationship to neuroimaging and CSF biomarkers. Alzheimers Dement. 2019;15: 232–244.
100. Copple BL, Li T. Pharmacology of bile acid receptors: Evolution of bile acids from simple detergents to complex signaling molecules. Pharmacol Res. 2016;104: 9–21.
101. Mertens KL, Kalsbeek A, Soeters MR, Eggink HM. Bile Acid Signaling Pathways from the Enterohepatic Circulation to the Central Nervous System. Front Neurosci. 2017;11: 617.
102. Quinn M, McMillin M, Galindo C, Frampton G, Pae HY, DeMorrow S. Bile acids permeabilize the blood brain barrier after bile duct ligation in rats via Rac1-dependent mechanisms. Dig Liver Dis. 2014;46: 527–534.
103. Xi J, Ding D, Zhu H, Wang R, Su F, Wu W, et al. Disturbed microbial ecology in Alzheimer's disease: evidence from the gut microbiota and fecal metabolome. BMC Microbiol. 2021;21: 226.
104. Identification of Faecalibacterium prausnitzii strains for gut microbiome-based intervention in Alzheimer's-type dementia. Cell Reports Medicine. 2021;2: 100398.
105. Lenoir M, Martín R, Torres-Maravilla E, Chadi S, González-Dávila P, Sokol H, et al. Butyrate mediates anti-inflammatory effects of Faecalibacterium prausnitzii in intestinal epithelial cells through Dact3. Gut Microbes. 2020;12: 1–16.
106. Jia L, Yang J, Zhu M, Pang Y, Wang Q, Wei Q, et al. A metabolite panel that differentiates Alzheimer's disease from other dementia types. Alzheimers Dement. 2022;18: 1345–1356.
107. Meoni G, Tenori L, Schade S, Licari C, Pirazzini C, Bacalini MG, et al. Metabolite and lipoprotein profiles reveal sex-related oxidative stress imbalance in de novo drug-naive Parkinson's disease patients. NPJ Parkinsons Dis. 2022;8: 14.
108. Tan AH, Chong CW, Lim S-Y, Yap IKS, Teh CSJ, Loke MF, et al. Gut Microbial Ecosystem in Parkinson Disease: New Clinicobiological Insights from Multi-Omics. Ann Neurol. 2021;89: 546–559.
109. Integrated Analyses of Microbiome and Longitudinal Metabolome Data Reveal Microbial-Host Interactions on Sulfur Metabolism in Parkinson's Disease. Cell Rep. 2019;29: 1767–1777.e8.
110. Yan Z, Yang F, Cao J, Ding W, Yan S, Shi W, et al. Alterations of gut microbiota and metabolome with Parkinson's disease. Microb Pathog. 2021;160: 105187.
111. Vascellari S, Palmas V, Melis M, Pisanu S, Cusano R, Uva P, et al. Gut Microbiota and Metabolome Alterations Associated with Parkinson's Disease. mSystems. 2020;5. doi:10.1128/mSystems.00561-20
112. Willkommen D, Lucio M, Moritz F, Forcisi S, Kanawati B, Smirnov KS, et al. Metabolomic investigations in cerebrospinal fluid of Parkinson's disease. PLoS One. 2018;13: e0208752.
113. Rosario D, Bidkhori G, Lee S, Bedarf J, Hildebrand F, Le Chatelier E, et al. Systematic analysis of gut microbiome reveals the role of bacterial folate and homocysteine metabolism in Parkinson's disease. Cell Rep. 2021;34: 108807.
114. Trivedi DK, Sinclair E, Xu Y, Sarkar D, Walton-Doyle C, Liscio C, et al. Discovery of Volatile Biomarkers of Parkinson's Disease from Sebum. ACS Cent Sci. 2019;5: 599–606.
115. Sender R, Fuchs S, Milo R. Revised Estimates for the Number of Human and Bacteria Cells in the Body. PLoS Biol. 2016;14: e1002533.
116. Qin J, Li R, Raes J, Arumugam M, Burgdorf KS, Manichanh C, et al. A human gut microbial gene catalogue established by metagenomic sequencing. Nature. 2010;464: 59–65.
117. Fischbach MA. Microbiome: Focus on Causation and Mechanism. Cell. 2018;174: 785–790.
118. Koh A, Bäckhed F. From Association to Causality: the Role of the Gut Microbiota and Its Functional

Products on Host Metabolism. Mol Cell. 2020;78: 584–596.
119. Kumar M, Ji B, Zengler K, Nielsen J. Modelling approaches for studying the microbiome. Nat Microbiol. 2019;4: 1253–1267.
120. Heinken A, Basile A, Hertel J, Thinnes C, Thiele I. Genome-Scale Metabolic Modeling of the Human Microbiome in the Era of Personalized Medicine. Annu Rev Microbiol. 2021;75: 199–222.
121. Blazier AS, Papin JA. Integration of expression data in genome-scale metabolic network reconstructions. Front Physiol. 2012;3: 299.
122. Hertel J, Heinken A, Martinelli F, Thiele I. Integration of constraint-based modeling with fecal metabolomics reveals large deleterious effects of spp. on community butyrate production. Gut Microbes. 2021;13: 1–23.
123. Cruz F, Faria JP, Rocha M, Rocha I, Dias O. A review of methods for the reconstruction and analysis of integrated genome-scale models of metabolism and regulation. Biochem Soc Trans. 2020;48: 1889–1903.
124. Baldini F, Heinken A, Heirendt L, Magnusdottir S, Fleming RMT, Thiele I. The Microbiome Modeling Toolbox: from microbial interactions to personalized microbial communities. Bioinformatics. 2019;35: 2332–2334.
125. Bauer E, Zimmermann J, Baldini F, Thiele I, Kaleta C. BacArena: Individual-based metabolic modeling of heterogeneous microbes in complex communities. PLoS Comput Biol. 2017;13: e1005544.
126. Thiele I, Sahoo S, Heinken A, Hertel J, Heirendt L, Aurich MK, et al. Personalized whole-body models integrate metabolism, physiology, and the gut microbiome. Mol Syst Biol. 2020;16: e8982.
127. Heinken A, Hertel J, Thiele I. Metabolic modelling reveals broad changes in gut microbial metabolism in inflammatory bowel disease patients with dysbiosis. NPJ Syst Biol Appl. 2021;7: 19.
128. Aden K, Rehman A, Waschina S, Pan W-H, Walker A, Lucio M, et al. Metabolic Functions of Gut Microbes Associate With Efficacy of Tumor Necrosis Factor Antagonists in Patients With Inflammatory Bowel Diseases. Gastroenterology. 2019;157: 1279–1292.e11.
129. Pryor R, Norvaisas P, Marinos G, Best L, Thingholm LB, Quintaneiro LM, et al. Host-Microbe-Drug-Nutrient Screen Identifies Bacterial Effectors of Metformin Therapy. Cell. 2019;178: 1299–1312.e29.
130. Baldini F, Hertel J, Sandt E, Thinnes CC, Neuberger-Castillo L, Pavelka L, et al. Parkinson's disease-associated alterations of the gut microbiome predict disease-relevant changes in metabolic functions. BMC Biol. 2020;18: 62.
131. Walker A, Schmitt-Kopplin P. The role of fecal sulfur metabolome in inflammatory bowel diseases. Int J Med Microbiol. 2021;311: 151513.
132. Matheus FC, Aguiar AS Jr, Castro AA, Villarinho JG, Ferreira J, Figueiredo CP, et al. Neuroprotective effects of agmatine in mice infused with a single intranasal administration of 1-methyl-4-phenyl-1,2,3,6-tetrahydropyridine (MPTP). Behav Brain Res. 2012;235: 263–272.
133. Song J, Hur BE, Bokara KK, Yang W, Cho HJ, Park KA, et al. Agmatine improves cognitive dysfunction and prevents cell death in a streptozotocin-induced Alzheimer rat model. Yonsei Med J. 2014;55: 689–699.
134. Campbell JM, Stephenson MD, de Courten B, Chapman I, Bellman SM, Aromataris E. Metformin Use Associated with Reduced Risk of Dementia in Patients with Diabetes: A Systematic Review and Meta-Analysis. J Alzheimers Dis. 2018;65: 1225–1236.
135. Patil SP, Jain PD, Ghumatkar PJ, Tambe R, Sathaye S. Neuroprotective effect of metformin in MPTP-induced Parkinson's disease in mice. Neuroscience. 2014;277: 747–754.
136. David LA, Maurice CF, Carmody RN, Gootenberg DB, Button JE, Wolfe BE, et al. Diet rapidly and reproducibly alters the human gut microbiome. Nature. 2014;505: 559–563.
137. Cassani E, Barichella M, Ferri V, Pinelli G, Iorio L, Bolliri C, et al. Dietary habits in Parkinson's disease: Adherence to Mediterranean diet. Parkinsonism Relat Disord. 2017;42: 40–46.
138. Ghosh TS, Rampelli S, Jeffery IB, Santoro A, Neto M, Capri M, et al. Mediterranean diet intervention alters the gut microbiome in older people reducing frailty and improving health status: the NU-AGE 1-year dietary intervention across five European countries. Gut. 2020;69: 1218–1228.
139. Smith PM, Howitt MR, Panikov N, Michaud M, Gallini CA, Bohlooly-Y M, et al. The microbial metabolites, short-chain fatty acids, regulate colonic Treg cell homeostasis. Science. 2013;341: 569–573.
140. Kujawska M, Jodynis-Liebert J. Polyphenols in Parkinson's Disease: A Systematic Review of In Vivo Studies. Nutrients. 2018;10. doi:10.3390/nu10050642
141. Ho L, Zhao D, Ono K, Ruan K, Mogno I, Tsuji M, et al. Heterogeneity in gut microbiota drive polyphenol metabolism that influences α-synuclein misfolding and toxicity. J Nutr Biochem. 2019;64: 170–181.
142. Mischley LK, Lau RC, Bennett RD. Role of Diet and Nutritional Supplements in Parkinson's Disease Progression. Oxid Med Cell Longev. 2017;2017: 6405278.
143. Magnusson KR, Hauck L, Jeffrey BM, Elias V, Humphrey A, Nath R, et al. Relationships between diet-related changes in the gut microbiome and cognitive flexibility. Neuroscience. 2015;300: 128–140.
144. Olson CA, Vuong HE, Yano JM, Liang QY, Nusbaum DJ, Hsiao EY. The Gut Microbiota Mediates the Anti-Seizure Effects of the Ketogenic Diet. Cell. 2018;173: 1728–1741.e13.
145. Boulos MI, Jairam T, Kendzerska T, Im J, Mekhael A, Murray BJ. Normal polysomnography parameters in healthy adults: a systematic review and meta-analysis. Lancet Respir Med. 2019;7: 533–543.
146. Yassine HN, Samieri C, Livingston G, Glass K, Wagner M, Tangney C, et al. Nutrition state of science and dementia prevention: recommendations of the Nutrition for Dementia Prevention Working Group. Lancet Healthy Longev. 2022;3: e501–e512.
147. Koeth RA, Wang Z, Levison BS, Buffa JA, Org E, Sheehy BT, et al. Intestinal microbiota metabolism of

L-carnitine, a nutrient in red meat, promotes atherosclerosis. Nat Med. 2013;19: 576.
148. Hirsch MA, Iyer SS, Sanjak M. Exercise-induced neuroplasticity in human Parkinson's disease: What is the evidence telling us? Parkinsonism Relat Disord. 2016;22 Suppl 1: S78–81.
149. Svensson M, Lexell J, Deierborg T. Effects of Physical Exercise on Neuroinflammation, Neuroplasticity, Neurodegeneration, and Behavior: What We Can Learn From Animal Models in Clinical Settings. Neurorehabil Neural Repair. 2015;29: 577–589.
150. Allen JM, Berg Miller ME, Pence BD, Whitlock K, Nehra V, Gaskins HR, et al. Voluntary and forced exercise differentially alters the gut microbiome in C57BL/6J mice. J Appl Physiol. 2015;118: 1059–1066.
151. Petriz BA, Castro AP, Almeida JA, Gomes CP, Fernandes GR, Kruger RH, et al. Exercise induction of gut microbiota modifications in obese, non-obese and hypertensive rats. BMC Genomics. 2014;15: 511.
152. Clarke SF, Murphy EF, O'Sullivan O, Lucey AJ, Humphreys M, Hogan A, et al. Exercise and associated dietary extremes impact on gut microbial diversity. Gut. 2014;63: 1913–1920.
153. Inamori T. [Experimental and clinical study of endolymphatic hydrops by means of electrocochleogram]. Nihon Jibiinkoka Gakkai Kaiho. 1984;87: 325–339.
154. Burokas A, Arboleya S, Moloney RD, Peterson VL, Murphy K, Clarke G, et al. Targeting the Microbiota-Gut-Brain Axis: Prebiotics Have Anxiolytic and Antidepressant-like Effects and Reverse the Impact of Chronic Stress in Mice. Biol Psychiatry. 2017;82: 472–487.
155. Azpiroz F, Dubray C, Bernalier-Donadille A, Cardot J-M, Accarino A, Serra J, et al. Effects of scFOS on the composition of fecal microbiota and anxiety in patients with irritable bowel syndrome: a randomized, double blind, placebo controlled study. Neurogastroenterol Motil. 2017;29. doi:10.1111/nmo.12911
156. Wang D, Ho L, Faith J, Ono K, Janle EM, Lachcik PJ, et al. Role of intestinal microbiota in the generation of polyphenol-derived phenolic acid mediated attenuation of Alzheimer's disease β-amyloid oligomerization. Mol Nutr Food Res. 2015;59: 1025–1040.
157. Barichella M, Pacchetti C, Bolliri C, Cassani E, Iorio L, Pusani C, et al. Probiotics and prebiotic fiber for constipation associated with Parkinson disease: An RCT. Neurology. 2016;87: 1274–1280.
158. Davari S, Talaei SA, Alaei H, Salami M. Probiotics treatment improves diabetes-induced impairment of synaptic activity and cognitive function: behavioral and electrophysiological proofs for microbiome-gut-brain axis. Neuroscience. 2013;240: 287–296.
159. Rao AV, Bested AC, Beaulne TM, Katzman MA, Iorio C, Berardi JM, et al. A randomized, double-blind, placebo-controlled pilot study of a probiotic in emotional symptoms of chronic fatigue syndrome. Gut Pathog. 2009;1: 6.
160. Akbari E, Asemi Z, Daneshvar Kakhaki R, Bahmani F, Kouchaki E, Tamtaji OR, et al. Effect of Probiotic Supplementation on Cognitive Function and Metabolic Status in Alzheimer's Disease: A Randomized, Double-Blind and Controlled Trial. Front Aging Neurosci. 2016;8: 256.
161. Tamtaji OR, Taghizadeh M, Daneshvar Kakhaki R, Kouchaki E, Bahmani F, Borzabadi S, et al. Clinical and metabolic response to probiotic administration in people with Parkinson's disease: A randomized, double-blind, placebo-controlled trial. Clin Nutr. 2019;38: 1031–1035.
162. Targeted microbiome-sparing antibiotics. Drug Discov Today. 2021;26: 2198–2203.
163. Park EM, Chelvanambi M, Bhutiani N, Kroemer G, Zitvogel L, Wargo JA. Targeting the gut and tumor microbiota in cancer. Nat Med. 2022;28: 690–703.
164. Mertsalmi TH, Pekkonen E, Scheperjans F. Antibiotic exposure and risk of Parkinson's disease in Finland: A nationwide case-control study. Mov Disord. 2020;35: 431–442.
165. Angelucci F, Cechova K, Amlerova J, Hort J. Antibiotics, gut microbiota, and Alzheimer's disease. J Neuroinflammation. 2019;16: 108.
166. Itzhaki RF. Herpes simplex virus type 1 and Alzheimer's disease: possible mechanisms and signposts. FASEB J. 2017;31: 3216–3226.
167. Chan FKL, To KF, Wu JCY, Yung MY, Leung WK, Kwok T, et al. Eradication of Helicobacter pylori and risk of peptic ulcers in patients starting long-term treatment with non-steroidal anti-inflammatory drugs: a randomised trial. Lancet. 2002;359: 9–13.
168. Efficacy and safety of ridinilazole compared with vancomycin for the treatment of Clostridium difficile infection: a phase 2, randomised, double-blind, active-controlled, non-inferiority study. Lancet Infect Dis. 2017;17: 735–744.
169. ZnO tetrapod materials for functional applications. Mater Today. 2018;21: 631–651.
170. Antoine TE, Hadigal SR, Yakoub AM, Mishra YK, Bhattacharya P, Haddad C, et al. Intravaginal Zinc Oxide Tetrapod Nanoparticles as Novel Immunoprotective Agents against Genital Herpes. J Immunol. 2016;196: 4566–4575.
171. Siebert L, Luna‐Cerón E, García‐Rivera LE, Oh J, Jang J, Rosas‐Gómez DA, et al. Light‐controlled growth factors release on tetrapodal ZnO‐incorporated 3D‐printed hydrogels for developing smart wound scaffold. Adv Funct Mater. 2021;31: 2007555.
172. Nasajpour A, Mandla S, Shree S, Mostafavi E, Sharifi R, Khalilpour A, et al. Nanostructured Fibrous Membranes with Rose Spike-Like Architecture. Nano Lett. 2017;17: 6235–6240.
173. Nasajpour A, Samandari M, Patil CD, Abolhassani R, Suryawanshi RK, Adelung R, et al. Nanoengineered Antiviral Fibrous Arrays with Rose-Thorn-Inspired Architectures. ACS Materials Letters. 2021. pp. 1566–1571. doi:10.1021/acsmaterialslett.1c00419
174. van Nood E, Vrieze A, Nieuwdorp M, Fuentes S, Zoetendal EG, de Vos WM, et al. Duodenal infusion of donor feces for recurrent Clostridium difficile. N Engl J Med. 2013;368: 407–415.
175. Bercik P, Denou E, Collins J, Jackson W, Lu J, Jury J, et al. The intestinal microbiota affect central levels

of brain-derived neurotropic factor and behavior in mice. Gastroenterology. 2011;141: 599–609, 609.e1–3.
176. Dimitriu PA, Boyce G, Samarakoon A, Hartmann M, Johnson P, Mohn WW. Temporal stability of the mouse gut microbiota in relation to innate and adaptive immunity. Environ Microbiol Rep. 2013;5: 200–210.
177. Kuai X-Y, Yao X-H, Xu L-J, Zhou Y-Q, Zhang L-P, Liu Y, et al. Evaluation of fecal microbiota transplantation in Parkinson's disease patients with constipation. Microb Cell Fact. 2021;20: 98.
178. Antunes LCM, Han J, Ferreira RBR, Lolić P, Borchers CH, Finlay BB. Effect of antibiotic treatment on the intestinal metabolome. Antimicrob Agents Chemother. 2011;55: 1494–1503.
179. Imhann F, Bonder MJ, Vich Vila A, Fu J, Mujagic Z, Vork L, et al. Proton pump inhibitors affect the gut microbiome. Gut. 2016;65: 740–748.
180. Bodrug SE, Warner BJ, Bath ML, Lindeman GJ, Harris AW, Adams JM. Cyclin D1 transgene impedes lymphocyte maturation and collaborates in lymphomagenesis with the myc gene. EMBO J. 1994;13: 2124–2130.
181. van Kessel SP, Frye AK, El-Gendy AO, Castejon M, Keshavarzian A, van Dijk G, et al. Gut bacterial tyrosine decarboxylases restrict levels of levodopa in the treatment of Parkinson's disease. Nat Commun. 2019;10: 310.
182. Jangi S, Gandhi R, Cox LM, Li N, von Glehn F, Yan R, et al. Alterations of the human gut microbiome in multiple sclerosis. Nat Commun. 2016;7: 12015.
183. Levi I, Gurevich M, Perlman G, Magalashvili D, Menascu S, Bar N, et al. Potential role of indolelactate and butyrate in multiple sclerosis revealed by integrated microbiome-metabolome analysis. Cell Rep Med. 2021;2: 100246.